# Structural and Transport Studies of Under Doped FeTe$_{1-x}$Se$_x$ (x=0.0, 0.01, 0.03, 0.05) Single Crystals

P.K. Maheshwari [1,2], Rajveer Jha[2], Bhasker Gahtori[2] and V.P.S. Awana[2,*]

[1] AcSIR-Academy of Scientific & Innovative Research- National Physical Laboratory, New Delhi-110012
[2] Quantum Phenomena and Applications, National Physical Laboratory (CSIR), New Delhi-110012, India
*Email: awana@mail.nplindia.org; web: awanavps.webs.com

**Abstract.** We report synthesis of under doped non-superconducting parent compound of iron chalcogenide, i.e FeTe$_{1-x}$Se$_x$ (x=0.0, 0.01, 0.03, 0.05) single crystals by flux free method. These FeTe$_{1-x}$Se$_x$ (x=0.0, 0.01, 0.03, 0.05) single crystals are crystallized in the tetragonal structure with space group P4/nmm at room temperature. Detailed scanning electron microscopy (SEM) results showed that the crystals are formed in slab-like morphology structure. Powder X-Ray diffraction (XRD) results show that (0 0 *l*) peaks are shifted towards higher angel with increasing Se concentration. Coupled magnetic and structural phase transition temperature, being seen as a step in resistivity, decreases from around 67K for x = 0.0 to 55K for x = 0.05 crystal. Further, the step in resistivity is seen to be hysteric (ΔT) in nature, i.e., different step temperatures in cooling and warming cycles. Interestingly, the hysteric (ΔT) becomes narrower from around 5.8K for x =0.0 to 0.95K for x =0.05 crystal.

The discovery of iron based superconductor attracted great interest of condensed matter physics scientific community since last couple of years or more. The Iron based superconductor having their superconducting critical temperature (T$_c$) up to 55 K, are different than the phonon-mediated conventional superconductors [1-4]. The parent component of Fe-based pnictide superconductors, i.e. REFeAsO possesses spin density wave (SDW) magnetic Fe. Similarly parent non superconducting compound of iron chalcogenide, i.e. FeTe possesses antiferromagnetically ordered Fe [5]. The parent compound of iron chalcogenide i.e. FeTe exhibits a drastic change at T~70 K in its electrical resistivity, magnetic susceptibility and heat capacity with decreasing temperature, which is related to first order phase (coupled magnetic and structural) transition [5,6].

In this short letter we report synthesis of under doped (low Se content) iron chalcogenide i.e., $FeTe_{1-x}Se_x$ (x=0.0, 0.01, 0.03, 0.05) single crystals without flux method. We used evacuated quartz tube without any complicated heating and cooling schedules related to travelling-solvent floating zone technique. The as synthesized $FeTe_{1-x}Se_x$ (x=0.0, 0.01, 0.03, 0.05) single crystals are crystallized in tetragonal structure with p4/nmm space group. We observed first order magnetic phase transition in electrical resistivity $\rho(T)$ measurements, which suppress with increasing Se doping concentration. The $\rho(T)$ measurements also show that during cooling and warming cycle there is considerable hysteresis ($\Delta T$) and that becomes narrow with increasing Se doping concentration

The single crystal of $FeTe_{1-x}Se_x$ (x=0.0, 0.01, 0.03, 0.05) were synthesized by self flux growth method [6,7]. The X-ray diffraction (XRD) measurements are done on a Rigaku X-Ray diffractometer using $CuK_\alpha$ line of 1.54184 Å. The morphology of the obtained FeTe single crystal is shown by scanning electron microscopy (SEM) images on a ZEISS-EVO MA-10 scanning electron microscope. Electrical resistivity measurements were performed on Quantum Design Physical Property Measurement System (PPMS) down to 2 K.

Figure 1 shows the X-ray diffraction analysis of crushed powder of $FeTe_{1-x}Se_x$ (x=0.0, 0.01, 0.03, 0.05) single crystals. These compounds are crystallized in the tetragonal structure with P4/nmm space group. Their X-ray diffraction analysis clearly shows that (0 0 *l*) peaks are shifted towards the higher angel, which specified that lattice parameter c (Å) decreases with increasing Se doping concentration. This result is verified through Reitveld refinement by using fullprof suite program. Namely, lattice parameters are a = 3.826(3), c = 6.292(2) for x = 0.0 decrease to a = 3.804(2), c = 6.241(2) for x = 0.05 compound.

Fig 2 shows the scanning electron microscopy (SEM) image of FeTe single crystal. This result shows the morphology of single crystals is having a slab like structure. EDX analysis showed that the compound is stoichiometric (Fe:Te= 1:1), see inset of Fig 2.

Figure 3 shows the temperature dependence of electrical resistivity ρ(T) for FeTe$_{1-x}$Se$_x$ (x=0.0, 0.01, 0.03, 0.05) single crystals in both cooling and warming cycle, without magnetic field from 300 K down to 4 K. This measurement shows that resistivity slightly increases with decreasing temperature down to coupled structural/magnetic phase transition, and after that the resistivity decreases with decreasing temperature for all the samples. Further, this measurement shows that phase transition temperature decreases with increasing Se doping concentration. Namely, the same decreases from around 67K for x =0.0 to 55K for x =.05 (5at% Se) crystal. From this measurement we can also conclude that resistivity increases slightly with increasing Se doping concentration.

Fig 4 shows the enlarged view of same ρ(T) curve in the temperature range 50-73 K. This curve shows that coupled magnetic/structural phase transition temperature is higher in warming than the cooling cycle. This phase transition temperature difference is defined by hysteresis (ΔT) which becomes narrow from 5.8 K to 0.95 K with increasing the Se doping concentration 0.0 to 0.05.

In conclusion, we have successfully synthesized single crystals of FeTe1-xSex (x=0.0, 0.01, 0.03, 0.05) through flux free technique. We observed that phase transition temperature decreases with increasing Se doping concentration. We also can conclude that the hysteresis (ΔT) of cooling and warming cycles becomes narrower with Se doping.

**FIGURE CAPTIONS**

**Figure 1**: The room temperature XRD pattern of crushed powder of FeTe$_{1-x}$Se$_x$ (x=0.0, 0.01, 0.03, 0.05) single crystal.

**Figure 2:** SEM image of FeTe single crystal at room temperature. Inset EDX analysis of FeTe single crystal

**Figure 3:** Temperature dependent electrical resistivity of FeTe$_{1-x}$Se$_x$ (x=0.0, 0.01, 0.03, 0.05) single crystals in cooling and warming cycles without magnetic field.

**Figure 4:** Zoomed view of the same ρ(T) curve in the temperature range 50-73 K.

*Fig. 1*

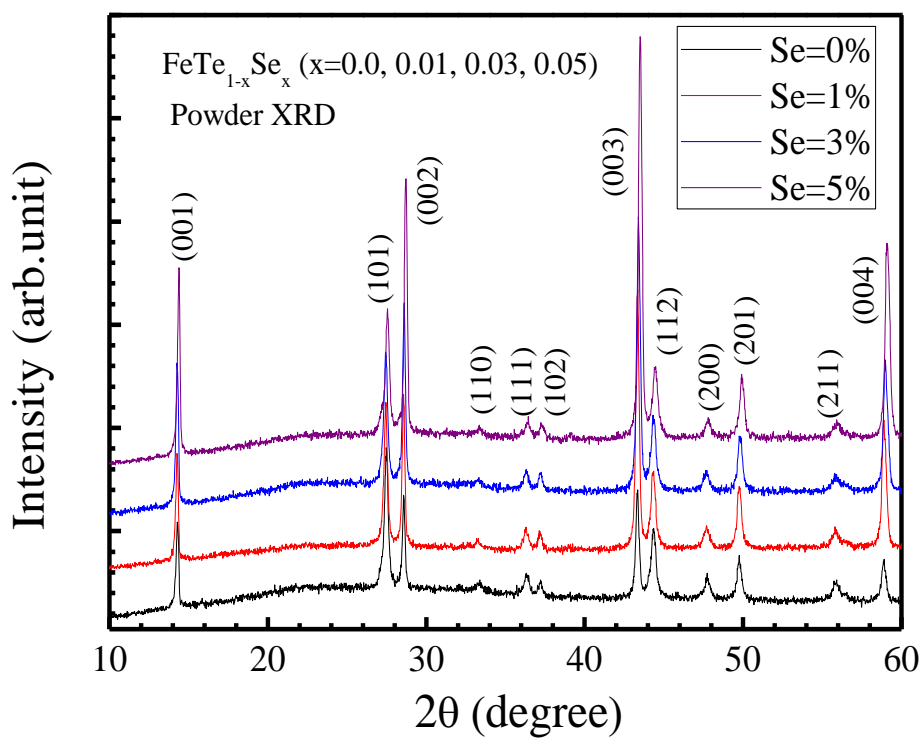

*Fig. 2*

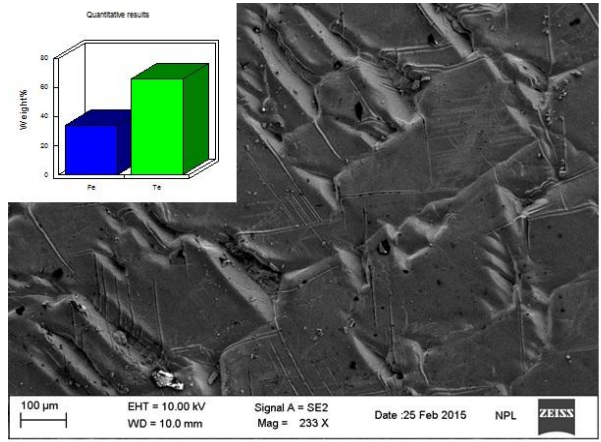

*Fig. 3*

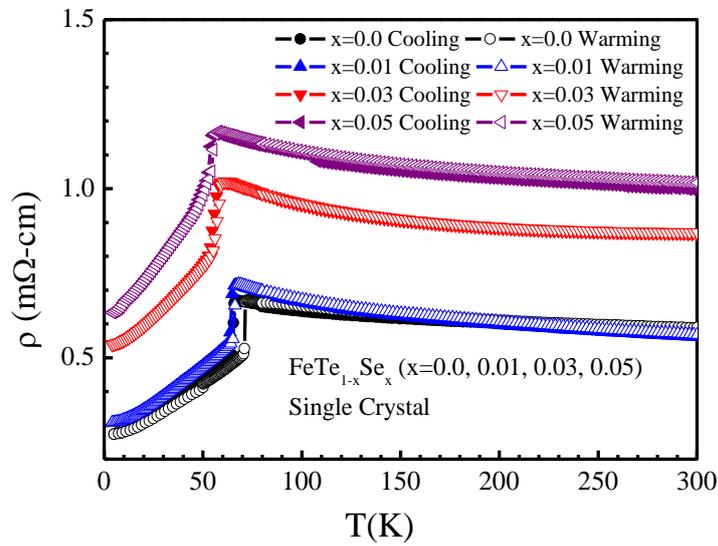

*Fig. 4*

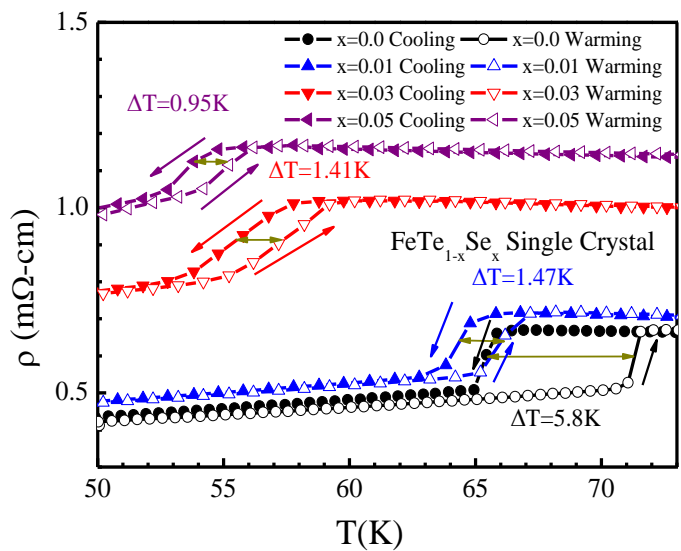